# Invicem Lorentz Oscillator Model (ILOM)

**Vladimir V. Semak[1] and Mikhail N. Shneider[2,a]**

1 Signature Science, LLC
2 Department of Mechanical and Aerospace Engineering, Princeton University

[a]email: m.n.shneider@gmail.com

*"Details sire the complications"*
*Old Chinese saying*

In this work we present detailed theoretical model of electron motion inside of an atom or molecule when subjected to electro-magnetic wave. This theoretical model, that we called Invicem Lorentz Oscillator Model (ILOM), contains Lorentz Oscillator Model (LOM) as a simplified approximation. The ILOM also contains what is called "nonlinear optics" as a second order approximation and it also demonstrates artificial nature of the separation of descriptions of electromagnetic wave propagation in media into Linear and Nonlinear Optics.

## I. INTRODUCTION

In all textbooks on electromagnetism and optics the interaction of electro-magnetic waves with matter is described using LOM [1]. Two crucial assumptions of this model are that the electron displaced from its equilibrium position inside of an atomic or molecular system by electric and magnetic fields of electro-magnetic wave will be subjected to 1) the returning force that is proportional to the electron displacement; and 2) the damping force that is proportional to the velocity of electron displacement. This model was formulated by Lorentz in early 1900s without any detailed explanations and justification of the foundations assumptions. A nice description of this model is given in book that presents lectures taught in Columbia University by Lorentz in the Spring of 1906 [1]. Since then, LOM demonstrated good results and became widely accepted without further questioning. In the meantime, a few inquisitive students asked for justification or explanation of; 1) why the returning force acting on electron is a linear function of the electron displacement form the equilibrium as if the electrons were attached to the nuclei by a spring; and 2) why the damping force is proportional to the electron speed as if there is some hydrodynamics like friction in the atom. While teaching laser physics classes for many years one of the authors received such questions every semester. The justifications for these, actually, not obvious assumptions were not provided neither by Lorentz [1] nor by all the followers ([2-4], for example).

Thus, the currently accepted equation of the aforementioned physical model is the equation of motion of a damped harmonic (linear) oscillator [1]:

$$\ddot{x} + \omega_0^2 x + \frac{\gamma}{m_e}\dot{x} = \frac{e}{m_e}E(t)\,, \tag{1}$$

where $\omega_0^2$ is the natural frequency of electron oscillations, $e$ and $m_e$ are charge and mass of electron, $\gamma$ is the damping factor, and $E(t)$ is the time variable external electric field (electric field of incident electro-magnetic wave, for example).

It is important to mention, that equation (1) is the foundational equation in the mathematical model used in the subset of physical optics. Physical effects related to the nonlinearity of motion of electrons in an atom exposed to variable, large amplitude electric field are not present in this equation. The nonlinearity of the electron motion and related to that physical effects are introduced by adding some terms in the follow up mathematical layouts. Thus, the nonlinearity of the electron motion and the effect of this nonlinearity on electro-magnetic wave propagation were never considered in a consistent manner since the LOM was published more than 100 years ago.

The second term in the equation of motion of a bound electron (equation 1) was included in the equation by Lorentz [1] in a "matter of fact" manner. All the following works and all the optics textbooks [2-4] followed Lorentz. Some authors provide somewhat in-depth consideration by suggesting that any potential of bonded electron described as a function, $U(x)$, of electron displacement from its equilibrium, $x$, can be approximated by Taylor series [2]:

$$U(x) = \frac{x^2}{2!}U^{(2)}(0) + \frac{x^3}{3!}U^{(3)}(0) + \frac{x^4}{4!}U^{(4)}(0) + \cdots, \qquad (2)$$

or

$$U(x) = ax^2 + bx^3 + cx^4 + \cdots. \qquad (3)$$

These considerations state that the first two terms in the Taylor series are absent because $U(0)=0$ for the equilibrium point $x=0$, and because the potential energy at this point has minimum, $U^{(1)}(0)=0$. Then, in case of small amplitude oscillations, $x \to 0$, one can limit expression for potential in the equation (2) to the quadratic term that leads to the second term in the equation (1).

If the amplitude, $x$, becomes large, the higher order terms must be taken into account and the returning force in the equation of motion becomes non-linear function of the electron displacement, $x$. Then, higher order terms are used in the Taylor series representing dependence of the electron potential on displacement, equation (2). It seems that it is common practice to leap over the detailed math derivations directly to the nonlinear polarization of medium that is dependent on electric field amplitude in the electro-magnetic wave (for example, [5]):

$$P = \varepsilon_0\chi^{(1)}E + \varepsilon_0\chi^{(2)}E^2 + \varepsilon_0\chi^{(3)}E^3 + \varepsilon_0\chi^{(4)}E^4 + \varepsilon_0\chi^{(5)}E^5 \ldots = P_l + P_{nl}, \qquad (4)$$

where $\varepsilon_0$ is the electric permittivity of free space, $\chi^{(1)}$ is the linear electric susceptibility and $\chi^{(2)}, \chi^{(3)}, \ldots, \chi^{(n)}, \ldots$ are, so called, nonlinear susceptibilities that, in general, are dependent on the electro-magnetic wave frequency (dispersion). From this expression for the polarization of a medium (equation (4)), it is easy to deduce an expression for the refractive index:

$$n(I) = n_0 + n_2 I + \cdots, \qquad (5)$$

where $n_0$ is the linear refractive index of medium. Obviously, such representation demands that, $n_0 \gg n_2 I \gg \cdots$. Indeed, for example for a quartz crystal $n_2 = 3\cdot 10^{-16}$ см$^2$/W.

This appears as obvious and satisfactory explanation for the second term in the left hand side of the equation (1); however, several seemingly uncomplicated questions tarnish clarity of this model. For example, how does the expression of the bound electron potential look in a form that is not simplified? What is the criterion for the oscillations to be "small" amplitude and what is the criterion for transition from “linear” to “nonlinear” optics? Also, perhaps, a methodology question – of whether it seems logical to solve “linear” equation and then manually “nonlinearize” the obtained dependencies by adding some “nonlinear” terms.

Similarly, the third term in the equation (1) representing the “damping force” that is proportional to the electron velocity [1] was introduced by Lorentz without any explanations. Such mathematical representation of the damping force is similar to the friction force that accompanies motion of a body in a medium. However, this similarity is obviously futile since electrons do not experience friction. It is interesting to note that Lorentz in his lectures taught at the Columbia University in 1906 and published one decade later [1] talks about aether. Although, nowhere in these lectures Lorentz explicitly relates the damping force to the electron friction, it is logical to suggest that this is what he had assumed. We know that aether doesn’t exist and it is peculiar that some text books still [5] denote this term as the “friction force”.

Exhaustive literature search reveals that only few publications [6, 7, 8, 9] discuss the nature of the proposed damping force. In 1907 it was suggested by Mandelstam [6] that an electron agitated by variable electric field of electro-magnetic wave moves with acceleration and, hence, it radiates. Since then, it became accepted [7, 8, 9] that the damping force originated from the electron energy loss due to this radiation. An interesting historic perspective related to the polemic between Rayleigh and Plank and the contribution of Lorentz to this polemic related to the nature of Raleigh scattering is given in the article [10] by Sobel’man. At this point it is necessary to clarify that, unlike the atomic orbital motion that also occurs with acceleration but doesn’t result in radiation obeying the restrictions imposed by the Bohr postulates, the electron “jitter” produces emission of radiation.

Thus, the electron displacement from the equilibrium atomic orbit induced by the external variable electric field produces dipole radiation. It has been shown that the damping force associated with the energy loss due to the dipole radiation is proportional to the third order time derivative of the electron displacement [6].

$$F_{rad} = \xi \dddot{r}\,, \qquad (6)$$

where $\xi = \dfrac{e^2}{6\pi\varepsilon_0 c^2}$.

It is common to assume that the motion of electron caused by the oscillating field of electro-magnetic wave can be approximated by harmonic oscillation with slowly varying amplitude. If that were true, then the third order time derivative in the equation (6) could be approximated by the first order time derivative, resulting in the Lorentz “friction” force, [8]. Indeed, for a slowly decaying periodic motion of an oscillator with a slowly varying amplitude of oscillations $a(t)$ $r(t) = a(t)\cos(\omega t)$, where $\dot{a}/a >> 2\pi/\omega$, the radiation force approximation is

$$F_{rad} = \xi \dddot{r} \approx -\xi\omega^2 \dot{r}\,. \qquad (7)$$

This expression of the damping force is commonly used in the model of the Lorentz oscillator in linear and nonlinear optics [2-4,8,9]. In this case, the "friction" coefficient is $\gamma = \xi\omega^2 / m$ .

It seems satisfactory to explain the damping force as related to the radiation “friction” that is proportional to the third order time derivative of the electron displacement from the equilibrium, equation (6). However, the simplification that assumes harmonic nature of forced electron oscillations and leads to "friction" force proportional to electron oscillation velocity, equation (7), appears to be not so straightforward.

First problem with such approximation is related to the nature of the frequency $\omega$ in the equation (7): should it be the natural oscillation frequency of electron, $\omega_0$, as in [8,9], or should it be frequency of the electro-magnetic wave, $\omega$, as in [11]. These frequencies may differ by many orders of magnitude and, hence, this question is of high practical importance.

Second problem is of the fundamental nature: is this a plausible assumption, that an electron motion induced due to interaction with electro-magnetic wave can be accurately approximated as harmonic oscillation? Our literature searches show that all publications lack the assessment of the range of electron displacement beyond which harmonic approximation becomes inaccurate. In this work we will show that, under laser interaction conditions that are typical for the realm of nonlinear optics, the harmonic oscillator assumption is inaccurate. However, even without complex theoretical consideration, the assumption of harmonic oscillations seems to fall a victim to a simple critique. Indeed, nonlinear optics covers the realm when the returning force in the equation (1) is nonlinear. Then, the electron motion is nonlinear and assuming that this motion as harmonic for the computation of dipole radiation related damping force represents a serious conflict.

In this work we will consider the missing steps and will attempt to offer deeper description of the response of matter to the variable electro-magnetic fields.

## II. FORMULATION OF INVICEM LORENTZ OSCILLATOR MODEL (ILOM)

The ILOM model formulation in most general form includes electron motion due to action of the force exerted by variable electric field of the incident electro-magnetic wave (we neglect force produced upon electron from the magnetic field of the wave) and the counteraction of the force returning the electron to its stable orbit and the force resisting electron movement:

$$m_e\ddot{r} + \frac{\partial U(r)}{\partial r} + F_{mr}(r) = eE(t), \tag{8}$$

where $r$ is the radial position of the electron, $U(r)$ is the potential in which electron moves around the nucleus, and $F_{mr}(r)$ is a motion resisting force.

The second term in the left hand side, expressed as the derivative of the electron potential over the radius of electron orbit, represents the returning force that acts on electron when it is displaced from the equilibrium orbit. We can denote this force as "displacement resistance force". If the electron potential is approximated as a quadratic function, than the returning force is a linear function as assumed in the Lorentz model.

The third term in the left hand side is another force that resists electron undulations caused by the electro-magnetic wave. We can denote this force as "motion resistance force". The first and obvious choice of the mechanism for this resistance is radiation related energy loss. The electron exposed to the electro-magnetic wave moves with acceleration and therefore it radiates and, thus, loses energy, hence, a "radiation friction" force acts on electron. As was mentioned above, this mechanism was proposed more than 100 years ago my Mandelstam [6]. Similarly to the authors of reference [5], we suggest that other, yet unexplored non-radiative mechanisms of electron energy loss are possible that produce motion resisting force. In our future works on further development of ILOM, we will investigate multi-body interaction mechanisms that involve cross-interaction of the electron undulating in the electro-magnetic wave with the other electrons and nucleus resulting in non-radiative electron energy loss.

Let us now modify ILOM equation of electron motion (8) by introducing electron displacement from the equilibrium position, $x = r - r_0$, where $r_0$ is the radius of equilibrium electron orbit. The equation motion will be as follows:

$$m_e\ddot{x} + \frac{\partial U(x+r_0)}{\partial x} + F_{mr}(x + r_0) = eE(t). \tag{9}$$

Further modification could be useful by introducing dimensionless variable $y = \frac{x}{r_0} = \frac{r-r_0}{r_0}$ and dividing both sides of the equation my mass of electron, $m_e$:

$$\ddot{y} + \frac{1}{m_e r_0^2}\frac{\partial U(yr_0+r_0)}{\partial y} + \frac{1}{m_e r_0}F_{mr}(yr_0 + r_0) = \frac{e}{m_e r_0}E(t)\ . \tag{10}$$

## III. RETURNING FORCE ACTING ON BOUND ELECTRON IN A REALISTIC POTENTIAL

Let us consider an example of possible displacement resistance force that acts on an electron in a hydrogen like atom. The effective potential of hydrogen like atom can be derived by considering counteracting forces of Coulomb attraction and centrifugal force acting on the electron in opposite direction. Here we will employ semi-classical approach.

The Coulomb force is

$$F_q = \frac{q_n e}{4\pi\varepsilon_0 r^2}\ , \tag{11}$$

and the centrifugal force is

$$F_c = m_e\omega_e^2 r\ \ , \tag{12}$$

where $e$ is the charge of electron, $q_n$ is some effective charge of the nucleus that acts on the electron, $r$ is the distance of electron orbit to the center of nucleus, $m_e$ is the mass of electron, and $\omega_\varepsilon$ is the electron rotation frequency. From the law of angular momentum conservation we have

$$m_e v_e r = m_e\omega_e r^2 = C\ , \tag{13}$$

where $v_e$ is the velocity of electron rotation, and $C$ is a constant that according to quantum mechanics postulate is proportional to a whole number times the reduced Planck's constant, $n\hbar$. From the equations (12) and (13) we have this expression of the centrifugal force

$$F_c = \frac{C^2}{m_e r^3}\ . \tag{14}$$

Finally, the resulting force acting on the electron is

$$F_\Sigma = F_q - F_c = \frac{q_n e}{4\pi\varepsilon_0 r^2} - \frac{C^2}{m_e r^3}, \tag{15}$$

where we assumed that the Coulomb force attracting electron to the nucleus is positive and the centrifugal force repelling electron from the nucleus is negative. Such radial distribution of force filed corresponds to the following potential

$$U(r) = \frac{C^2}{2m_e r^2} - \frac{q_n e}{4\pi\varepsilon_0 r} = \frac{a}{r^2} - \frac{b}{r}\ . \tag{16}$$

We can further modify this formula expressing constants $a$ and $b$ in terms of ionization potential $U_0$, and electron orbit radius, $r_0$. The resulting force acting on the electron rotating on its orbit should be zero when the electron is exactly on the orbit

$$F(r = r_0) = \left.\frac{dU}{dr}\right|_{r=r_0} = -\frac{2a}{r_0^3} + \frac{b}{r_0^2} = 0 . \tag{17}$$

Then $\frac{2a}{r_0} = b$ and equation (17) takes form containing only one parameter, a,

$$U(r) = a\left(\frac{1}{r^2} - \frac{2}{r_0 r}\right) . \tag{18}$$

The electron is at its minimum potential that equals to negative value of the potential of ionization, - $U_0$, when electron is at its exact orbit, i.e. $r = r_0$. Then,

$$U|_{r=r_0} = a\left(\frac{1}{r_0^2} - \frac{2}{r_0^2}\right) = -U_0 , \tag{19}$$

and, therefore, $a = U_0 r_0^2$. Finally, from the equations (17) and (18) we have equations for the electron potential

$$U(r) = U_0 r_0^2 \left(\frac{1}{r^2} - \frac{2}{r_0 r}\right) , \tag{20}$$

and the returning force acting on electron at the orbit

$$F(r) = \frac{dU}{dr} = 2U_0 r_0^2 \left(\frac{1}{r_0 r^2} - \frac{1}{r^3}\right). \tag{21}$$

It is important to note that comparing equations (15) and (21) leads to conclusion that the following equalities must be true: $2U_0 r_0 = \frac{q_n e}{4\pi\varepsilon_0}$ and $2U_0 r_0^2 = \frac{C^2}{m_e}$. Taking as an example atom of hydrogen and substituting corresponding values, i.e. equating $U_0$ to hydrogen ionization potential of 13.9 eV, equating $r_0$ to the Bohr's radius, and equating constant $C$ to the reduced Planck's constant $\hbar$, and also assuming that the effective charge of nucleus equals to the absolute charge of the electron ($q_n$=e), one can find that the above equalities are satisfied with reasonable accuracy.

## IV. DAMPING FORCE CORRESPONDING TO ELECTRON ENERGY LOSS

As we described above, one of the possible damping force physical mechanisms is the radiative energy loss of oscillating electron. The radiation energy losses correspond to the "radiation friction" force (the Abraham-Lorentz force [9,11,12]) that acts on the oscillating electron:

$$\vec{F}_{rad} = \frac{e^2}{6\pi\varepsilon_0 c^2}\dddot{x} = \xi\,\dddot{x} , \ \xi = \frac{e^2}{6\pi\varepsilon_0 c^2} . \tag{22}$$

No further approximations and simplifications will be used in our model (ILOM).

## V. ILOM EQUATION OF ELECTRON MOTION FOR HYDROGEN LIKE ATOM

Now let us use equation for returning force (21) corresponding to a realistic electron potential (20) in order to deduce exact equation of motion of electron subjected to the oscillating electric field of electro-magnetic wave. Using expression for the returning force given by the equation (13), the equation for electron motion becomes

$$\ddot{r}+\frac{2U_0r_0^2}{m_e}\left(\frac{1}{r_0r^2}-\frac{1}{r^3}\right)-\frac{\xi}{m_e}\dddot{r}=\frac{e}{m_e}E, \tag{23}$$

where $\xi=\frac{e^2}{6\pi\varepsilon_0c^3}$.

Next we will modify this equation by expressing electron motion in terms of its displacement from the equilibrium orbit, $x$, similar to the equation (1). For this purpose we will substitute variables in the equation (23) introducing $x = r - r_0$

$$\ddot{x}+\frac{2U_0r_0^2}{m_e}\left(\frac{1}{r_0(x+r_0)^2}-\frac{1}{(x+r_0)^3}\right)-\frac{\xi}{m_e}\dddot{x}=\frac{e}{m_e}E. \tag{24}$$

This is complete electron motion equation that constitutes ILOM and, as we will show below, it produces solution that differs from the approximate all-familiar Lorentz LOM equation (1).

At this point it is worth noting that the equation (20) was deduced for the electron potential model that we have described above. This model should be rather adequate for simple atoms, like hydrogen and helium. One can modify and improve this model to have realistic and more accurate form of the electron potential for complex atoms and molecules. The latter would require use of quantum mechanical models.

Let's modify equation of motion (24) by introducing dimensionless variable $y=x/r_0$ where $r_0$ is the radius of electron orbit, and x is the electron displacement from the orbit. Then, the equation for dimensionless displacement of electron relative to its stable orbit will have form

$$\ddot{y}+\frac{2U_0}{m_er_0^2}\left(\frac{1}{(y+1)^2}-\frac{1}{(y+1)^3}\right)-\frac{\xi}{m_e}\dddot{y}=\ddot{y}+\frac{2U_0}{m_er_0^2}\left(\frac{y}{(y+1)^3}\right)-\frac{\xi}{m_e}\dddot{y}=\frac{e}{m_er_0}E, \tag{25}$$

where $\xi=\frac{e^2}{6\pi\varepsilon_0c^3}$.

## VI. ANALYSIS OF RETURNING FORCE IN ILOM ELECTRON MOTION EQUATION

Since the nonlinearity is dictated by the deviation of returning force from linear function let's explore the behavior of the function for hydrogen like atom.

For a hydrogen like atom the term in parentheses in the left-hand side of the equation (25) that represents returning force divided by electron mass and electron orbit radius is plotted as a function of electron displacement divided by the electron orbit radius in the Figure 1 a) and b). Here, of course, negative and positive values of dimensionless displacement correspond to motion toward and away from the nucleus, correspondingly.

It is easy to see that, for the displacement of electron that are on the order of ~$0.1r_0$ the returning force is strongly asymmetric and differs significantly from linear function. This corresponds to highly nonlinear response of electron motion to the oscillating electro-magnetic wave. However, even for relative displacement of ~0.01 the asymmetry and nonlinearity of the returning force is notable; however, it is relatively much smaller (Figure 1.b).

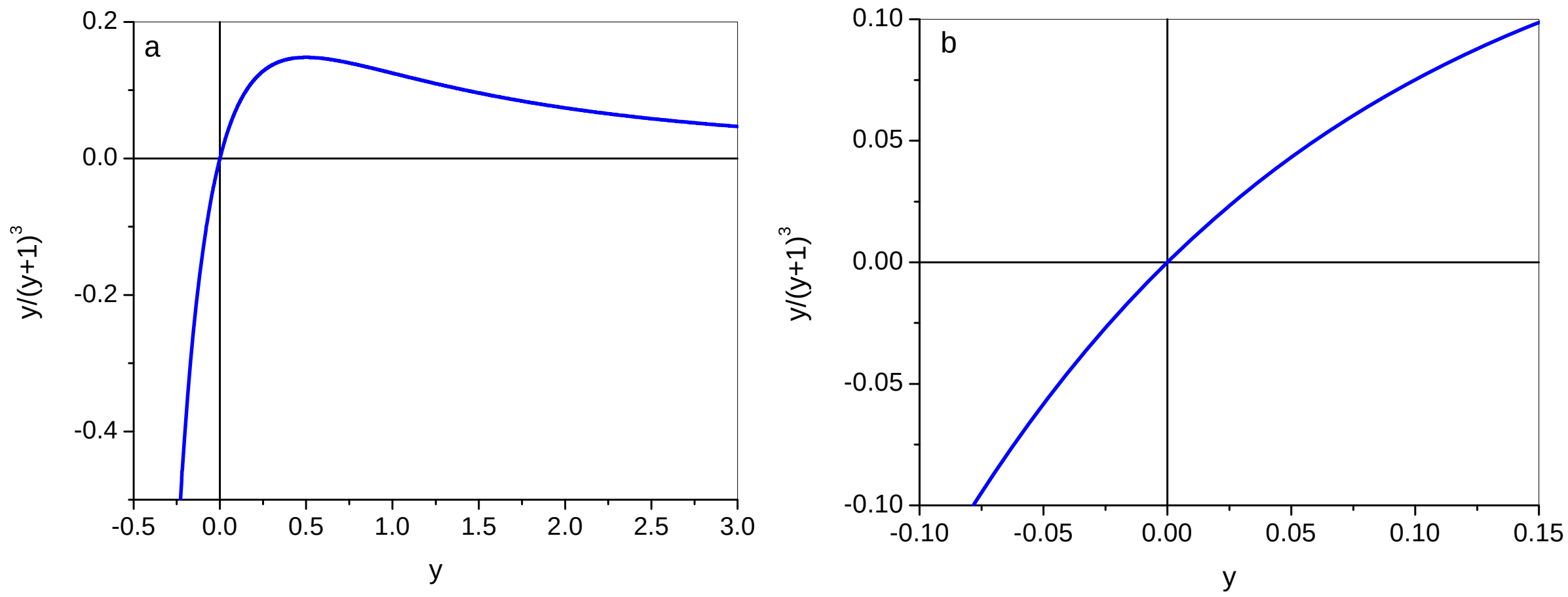

Fig. 1. a), b). Returning force normalized to electron mass and electron orbit radius (vertical axis normalized to $\frac{2U_0}{m_e r_0^2}$), plotted as a function of electron displacement from the orbit normalized to the electron orbit radius (horizontal axis – $y=(r-r_0)/r_0$) for a hydrogen like atom. Portions of the same graph are shown in a) and b) with different magnification. Negative $y$ is toward the nucleus and positive is away.

Similar observations are for the electron potential, shown as function of dimensionless electron displacement (Figure 2. a, b)

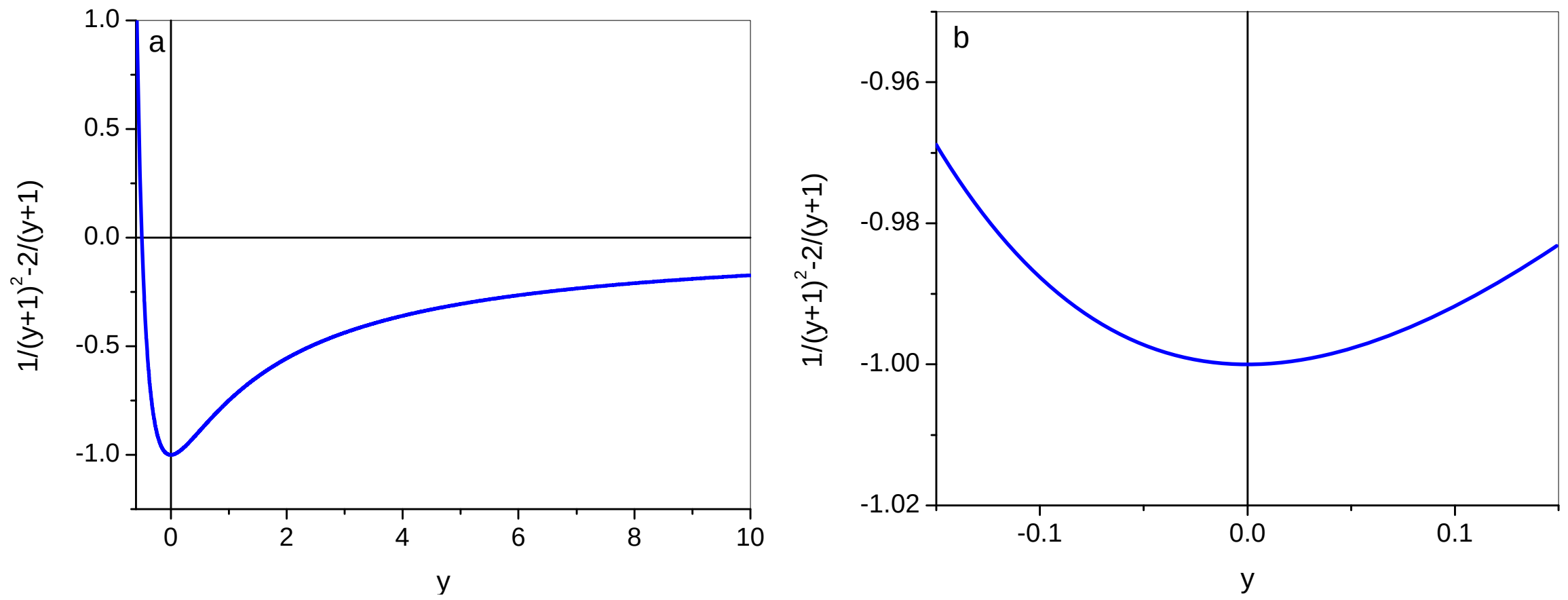

Fig. 2. a), b). Potential of the electron normalized to the ionization potential (vertical axis normalized to $U(r)/U_0$) plotted as a function of electron displacement from the orbit normalized to the electron orbit radius (horizontal axis - $y=(r-r_0)/r_0$) for a hydrogen like atom. Portions of the same graph are shown in a) and b) with different magnification. Same as above, negative $y$ is toward the nucleus and positive is away.

Let us now consider representing term in parenthesis in the left-hand side of the equation (25) as a Taylor series approximation. The Tylor series representation of the third term is as follows

$$\frac{y}{(y+1)^3} = \sum_{n=2}^{\infty} \frac{1}{2}(-1)^n (n-1) n y^{n-1} = y - 3y^2 + 6y^3 - \cdots. \quad (26)$$

Retaining first three terms of the Taylor series gives following equation of electron motion

$$\ddot{y} + \frac{2U_0}{m_e r_0^2}(y - 3y^2 + 6y^3) - \frac{\xi}{m_e}\dddot{y} = \frac{e}{m_e r_0} E. \quad (27)$$

Usually, textbooks present similar equation of motion. Of course, since detailed consideration is not performed in these books, the third term in the left-hand side of the equation (27) is given as $ay + by^2 + cy^3$ with some unknown coefficients $a$, $b$, and $c$ ([2,5], for example). In these textbooks a comment is made that "symmetric systems" are approximated by sum of linear and quadratic terms, i.e. $a, b \neq 0; c = 0$, while "asymmetric systems" are approximated by linear and cubic terms, i.e. $a, c \neq 0; b = 0$. Our above consideration, that provides students with in depth understanding of the subject, demonstrates, in particular, groundless nature of statements like this. It clearly follows from our consideration that the electron always resides in an asymmetric potential and, consequently, the returning force is always nonlinear and asymmetric. The approximation of these potential or force by a Taylor series is just a mathematical procedure that does not carry any physical significance.

One additional observation can be also of an educational importance for students: optics is always nonlinear and subdivision of optics into linear and nonlinear is dictated by the mathematical convenience and desired accuracy of the description of electron motion. Thus, unlike, for example, division into classical and quantum mechanics, division of optics into linear and nonlinear is not representative of difference in the physical model.

From the Figures 1 and 2, one can qualitatively see that the deviation of returning force from linear function becomes noticeable by eye for the amplitudes exceeding approximately $0.01r_0$. Let's now estimate the laser parameters for which electron motion "becomes" nonlinear. For this purpose we will use equation (1) in which returning force approximated as a linear function and the electron motion is assumed to be a harmonic function. Thus, we will use well known formula for the amplitude of electron oscillations

$$x_0 = \frac{e/m_e}{\omega^2 - \omega_0^2 + i\gamma\omega} E_0. \quad (28)$$

Let us take, for example, hydrogen atom with ionization potential $U_0$=13.6 eV. The natural oscillation frequency then is $\omega_0 = 4.13 \cdot 10^{16}$ rad/s. Assuming laser pulse energy 0.1J, pulse duration 10ns, wavelength of 1.06 μm ($\omega = 2\pi c/\lambda$), focused in a spot with radius 50 μm on 1/e intensity level, the amplitude of electric field is $E_0 = 9.78 \cdot 10^7$ V/m Substituting these values in the equation (28) gives the amplitude of electron oscillations $x_0 \sim 10^{-13}$ m. Dividing this amplitude by the value of Bohr's radius $r_0 = 5\ 10^{-11}$m gives relative amplitude of electron oscillations $y_0 \sim 10^{-2}$.

We should mention that, in real case scenario one can formulate a quantitative criterion that provides ranges of electron displacement where linear approximation delivers acceptable accuracy. It is reasonable to expect that such criteria will strongly depend on the physical conditions. For example, in case of low intensity light propagation at distances of several meters the linear approximation can provide accuracy that is much better than needed in practice. However, the same intensity light propagating at distance of billions lightyears can accumulate effects related to the nonlinearity of the motion of electrons in the atoms of interstellar and intergalactic matter, such that consideration that includes motion of electron in real potential will be required for acceptable accuracy. Same can be suggested for the case of laser beam propagation inside of the resonators with extremely high Q-factor.

## VII. NUMERICAL SOLUTION OF ILOM EQUATION OF MOTION AND DISCUSSIONS

A nonlinear second-order ordinary differential equation (25) was represented as a Cauchy problem for a system of two first-order differential equations

$$\dot{y} = z$$

$$\dot{z} = \frac{e}{mr_0} E_a \cos(\omega t) - \frac{2U_0}{mr_0^2} \frac{y}{(y+1)^3} + \frac{\xi}{m} \dddot{y} \tag{29}$$

with the initial conditions

$$y(0) = 0, \quad z(0) = \dot{y}(0) = 0. \tag{30}$$

The system of ordinary differential equations (29) with the initial conditions (30) was solved by the standard Runge-Kutta method of order 4 accuracy [13]. The term expressing the damping force $\propto \dddot{y}$ was taken from the previous time step, thereby allowing the problem to be solved by an explicit method.

The computation results demonstrating transition from "linear" to "nonlinear" response of the electron in laser field under typical parameters of a nanosecond pulse Nd:YAG laser are shown in the Figure 3. The computations were performed for an imaginary hydrogen like atom with the low ionization potential – 1 eV. The phase diagrams shown in the Figure 3 show gradual transition of the phase diagram from circular or elliptical for the low laser pulse energies( 0.1mJ and 1mJ) to the once that exhibit some degree of nonlinearity (5mJ and 10mJ) to the once that show very high nonlinearity of the of the electron response (50mJ and 100mJ). The ionization potential of hydrogen is 13.5984 eV and, therefore, the transition to noticeably nonlinear (nonharmonic) electron motion will take place at higher laser pulse energies. However, the considered conditions demonstrate possible response of a Rydberg atom.

The effect of variation of the electron potential depth is shown in Figure 4 for a 50mJ, 10ns pulse of a Nd:YAG laser focused in a spot with 20µm radius. Under the considered interaction conditions, the transition from linear (harmonic) to nonlinear electron response occurs for the hypothetical atoms with the ionization potential below 5eV.

Perhaps the most interesting result of simulation using ILOM is shown in the Figure 5 (a-c). The red line shows the electron response to a femtosecond and nanosecond laser pulses shown with a blue line. The top graph shows the temporal position of the electron of a hydrogen like atom with the ionization potential of 1eV exposed to the pulse of second harmonic of Nd:YAG laser with the full width at half maximum of 20fs and beam intensity of 3.975 $10^{15}$W/m$^2$ (Figure 5. a) The simulation shows that electron exposed to ultrashort pulse of relatively high intensity undulates at two characteristic frequencies – frequency of the incident light, ω, and at the natural oscillation frequency, $\omega_0$. This means that "pinning" an atom with fast front and high intensity laser pulse will produce scattering at the central frequency that equals laser frequency (Rayleigh scattering) and the emission at different central frequency that equals eigenfriquency of the electron oscillations in the atom. Decrease of the raise/fall rate of the laser intensity by increasing the laser pulse duration (Figure 5. b) of decreasing the maximum laser intensity (Figure 5. c) results in electron undulations at only one frequency - frequency of laser. Here, we should mention that this effect was known for mechanical or electric oscillators [14]; however, in optics the nonstationary (transitional) behavior of the electron oscillations was never consider prior to our work.

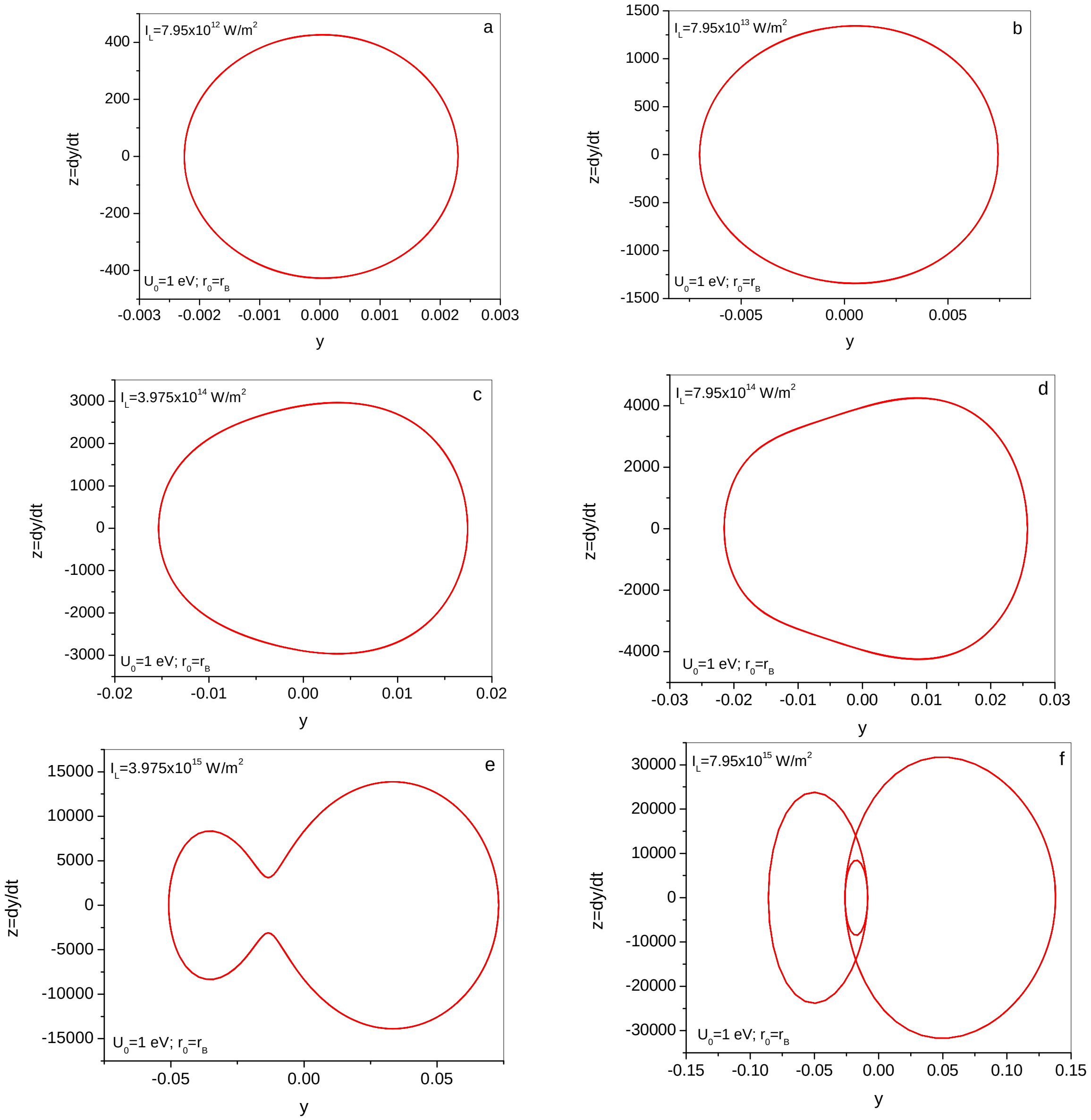

Fig. 3 (a-f). Calculation results showing the effect of the variation of the pulse energy of 10ns laser focused in a spot with radius 20 microns computed for a hydrogen like atom with ionization potential $U_0$=1eV and electron orbit radius that is equal to the Bohr's radius, $r_0=r_b$

a – 0.1 mJ; b - 1mJ; c- 5mJ; d - 10mJ; e – 50mJ; f – 100mJ

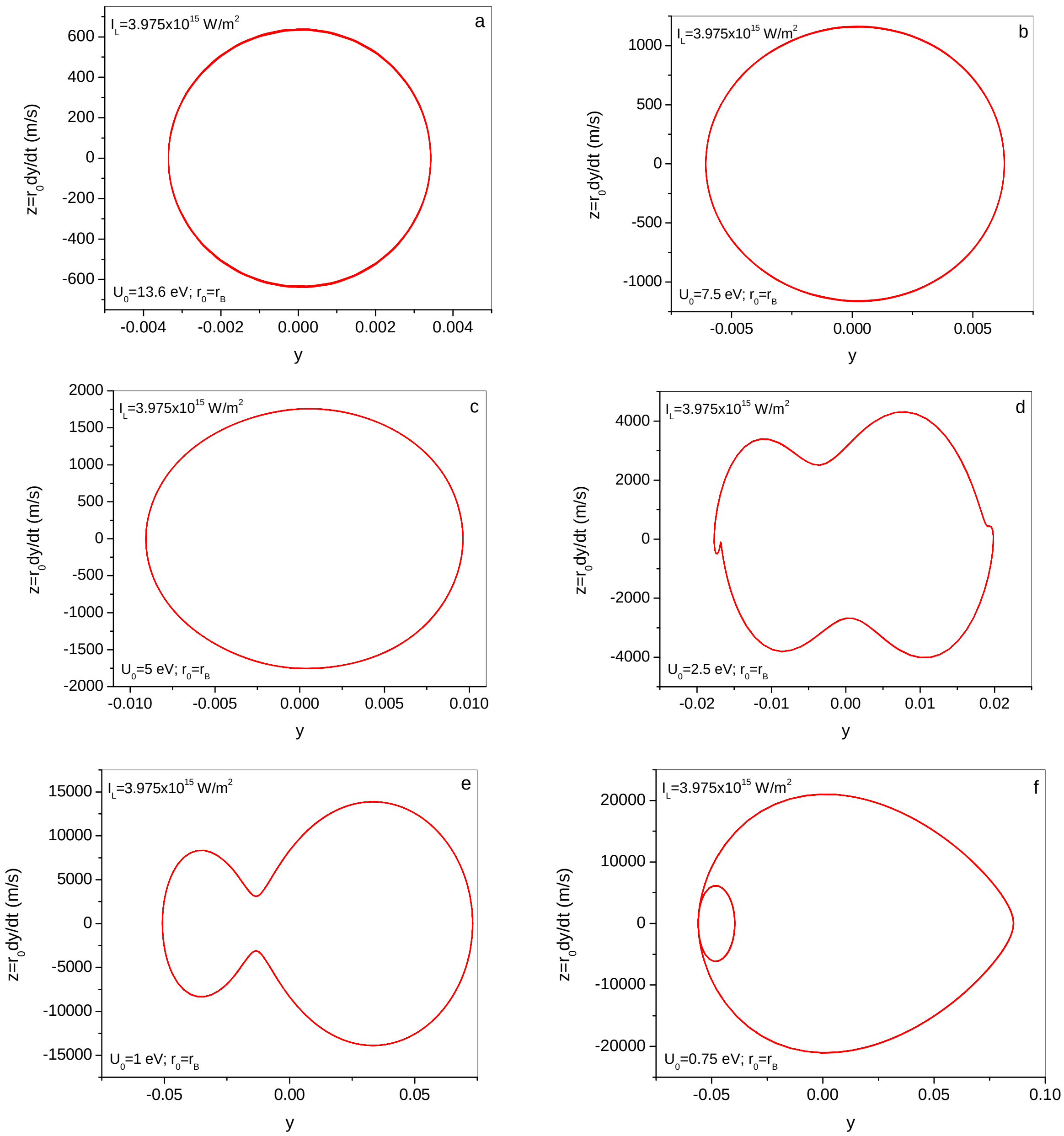

Fig. 4 (a-f). Calculations showing the effect of variation of the electron potential $U(r)$ of a hydrogen like atom (left lower corner) for a 50 mJ 10ns laser pulse focused in a spot with radius 20 microns. The electron orbit radius equals to the Bohr's radius.

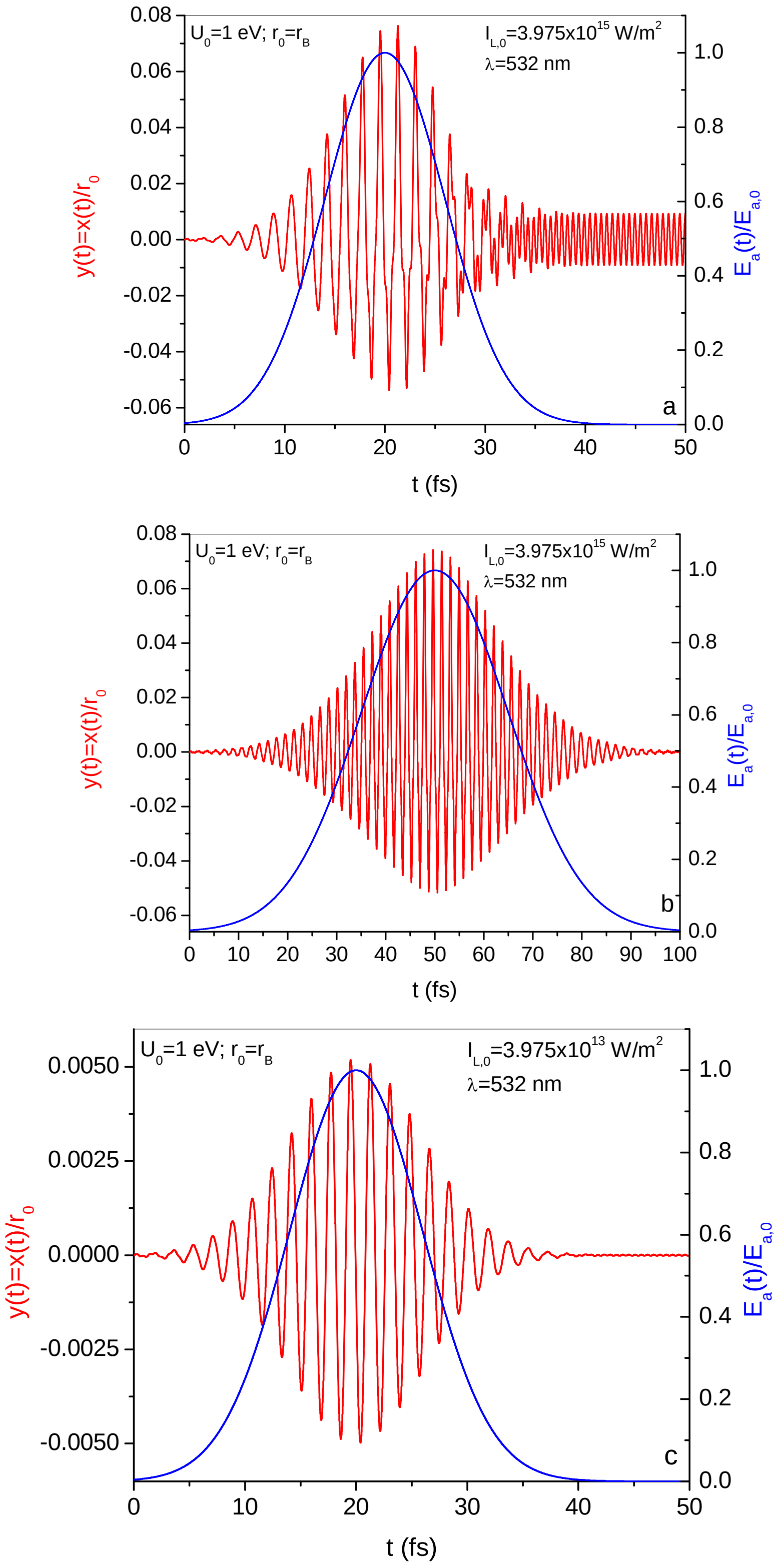

Fig. 5. Calculations showing the effect of "pinging" an atom with laser pulses of variable duration, $\tau_p$ and intensity, $I_0$. Computations performed for hydrogen like atom with ionization

potential $U_0$= 1eV, and electron orbit radius that equals Bohr's radius, $r_0$=$r_B$. a) $\tau_p$ =15fs , $I_0$=3.975 $10^{15}$W/m$^2$; b) $\tau_p$ =40fs , $I_0$=3.975 $10^{15}$W/m$^2$; c) $\tau_p$ =20fs , $I_0$=3.975 $10^{13}$W/m$^2$;

## VIII. CONCLUSIONS

The new model called Instead Lorentz Oscillator Model (ILOM) provides more adequate and logical description of response of a bound electron to incident electro-magnetic wave since it is based on the fundamental principles instead of "simple" and "obvious" assumptions.

The developed new model treats interaction of electro-magnetic wave with matter in a unified manner, In particular, our model demonstrates that separation of the realm of physical optics into "linear" and "nonlinear" is artificial and serves only purpose of mathematical convenience.

The ILOM provides theoretical foundation for detailed and consistent deduction of the formulae that express the refractive index of materials as function of the electro-magnetic wave and the fundamental material properties. The ILOM includes current "linear" and "nonlinear" optics as subsets and also opens the new subset of transitional linear-to-nonlinear optics. In particular, it seemingly follows from the estimates that the interaction with most of the short duration moderate-to-high energy laser pulses falls into this transitional region. Practical application of ILOM allows to quantitatively ascertain the level of mathematical complexity that is required in order to achieve needed accuracy of the simulation of the effects of electro-magnetic wave interaction with matter.

The ILOM opens an opportunity of designing materials with extreme nonlinearity of the real and imaginary parts of the refractive index. For example, our preliminary calculations using ILOM demonstrate that Rydberg atoms will exhibit relatively high nonlinear response.

For the first time since inception of this field of optics, we included in consideration nonstationary response of a bound electron to an electro-magnetic wave. The simulations demonstrated an amazing prediction that, pining atom or molecule with a pulse with sharp raise/fall front will produce commonly known Rayleigh scattering in combination with the emission with central frequency that equals to the frequency of the natural oscillations of bound electron. This result should be experimentally verified and it can have significant practical applications, for example, in stand-off detection and LDAR atmospheric studies.

## ACKNOWLEDGMENTS

One of the authors (VS) would like to thank Signature Science. LLC for providing tremendous support for this research activity.